\documentclass{ws-procs9x6}
\usepackage{amsmath}
\newcommand{\DI}{\Delta I}

\begin{document}

\title{NUCLEAR $\beta$ DECAY WITH LORENTZ VIOLATION}

\author{J.P.\ NOORDMANS,$^*$ H.W. WILSCHUT, and R.G.E.\ TIMMERMANS}

\address{KVI, University of Groningen\\
Zernikelaan 25, NL-9747 AA Groningen, The Netherlands\\
$^*$E-mail: J.P.Noordmans@rug.nl}

\begin{abstract}
We consider the possibility of Lorentz-invariance violation in weak-decay processes. We present a general approach that entails modifying the $W$-boson propagator by adding a Lorentz-violating tensor to it. We describe the effects of Lorentz violation on nuclear $\beta$ decay in this scenario. In particular we show the expression for a first-forbidden transition with a spin change of two. Using data from an old experiment on the rotational invariance of yttrium-90, we derive several bounds on the Lorentz-violating parameters of the order of $10^{-6}$-$10^{-8}$.
\end{abstract}

\bodymatter

\section{Introduction}
There has been considerable scientific interest in the possibility of the violation of Lorentz symmetry, in the context of searches for a theory of quantum gravity. Its phenomenological consequences have been studied extensively, in particular in the context of the Standard-Model Extension (SME)\cite{sme}. Extensive experimental efforts have been made to bound the coefficients in all sectors of the SME\cite{datatables}. 

Missing in these efforts is the study of weak decay. Until recently, there were, to our knowledge, only two dedicated experiments that addressed rotational invariance violation in $\beta$ decay\cite{newman, ullman}. This effort is now supplemented by an experiment at KVI\cite{kviexp}. 

Theoretically, the issue of calculating cross sections and decay rates is also far from settled, although some amount of work has already gone into this\cite{crossdecay}. 

We started a joint experimental and theoretical effort at KVI, addressing the issue of Lorentz violation in nuclear $\beta$ decay. We incorporate Lorentz violation in weak decay by modifying the gauge boson propagator to
\begin{equation}
\left\langle W^{\mu+}(p)W^{\nu-}(-p)\right\rangle = \frac{-i(g^{\mu\nu}+\chi^{\mu\nu})}{M_W^2}\ ,
\label{Wpropagator}
\end{equation}
where $g^{\mu\nu}$ is the usual Minkowski metric and $\chi^{\mu\nu}$ a general complex (possibly momentum-dependent) tensor, that parametrizes the Lorentz violation. What we miss in this way are mainly kinematic properties of the external particles (see Ref.\ \refcite{lvbetadecay1} for more discussion).

\section{$\beta$-decay transitions and forbiddenness}
$\beta$-decay transitions are classified by the spin change from parent to daughter  nucleus ($\DI$) and the parities of the parent and daughter nuclei, denoted by $\pi_i$ and $\pi_f$, respectively. 

One distinguishes different $\beta$-decay transitions also by their {\it forbiddenness}, or the degree of suppression of the quantum mechanical amplitude that describes the process. This suppression is determined by factors of three small quantities: $R/\lambda$, $v_N$ and $\alpha Z$, which are the ratio of the nuclear radius and the de Broglie wavelength of the leptons, the velocity of the decaying nucleon in units of $c$, and the fine-structure constant times the charge of the daughter nucleus, respectively. Factors of $R/\lambda$  come from the leptons carrying away orbital angular momentum (corresponding to higher order terms in the multipole expansion of the lepton wavefunctions). The quantities $v_N$ and $\alpha Z$ originate from relativistic and Coulomb effects, respectively.

If we assume these three quantities to vanish, we limit ourselves to so-called {\it allowed} $\beta$-decay transitions. The leptons do not carry away orbital angular momentum in this case. Consequently, the total angular momentum of the lepton pair can be either zero or one. In the Lorentz-symmetric case this translates directly into the selection rules for allowed $\beta$ decay, which can have spin change zero or one while the relative parity must be $\pi_i \pi_f = +1$. With Lorentz violation this connection is not that direct anymore, as explained later.

When allowing for factors of $R/\lambda$, $v_N$ and $\alpha Z$, the leptons can carry off orbital angular momentum and the total angular momentum of the lepton pair can be $J=0,1,2,\ldots$, which allows for a larger spin change in the transition, as well as the possibility of a relative parity of $\pi_i\pi_f = -1$. For a complete classification of transitions and their forbiddenness in the Lorentz-symmetric and Lorentz-violating cases, see Ref.\ \refcite{lvbetadecay3}.

\section{The effects of Lorentz violation}
Using the propagator in Eq.~\eqref{Wpropagator}, we have calculated the rate of general $\beta$-decay transitions. The expression for allowed $\beta$ decay can be found in Ref.~\refcite{lvbetadecay1}, and the one for forbidden transitions can be found in Ref.~\refcite{lvbetadecay2}.

In Refs.\ \refcite{lvbetadecay3} and \refcite{lvbetadecay2} we found that there is a relative enhancement of Lorentz-violating effects in transitions with $\DI \geq 2$ by a factor of $\alpha Z / pR$. From the very schematic expression
\begin{equation}
(1,\boldsymbol{\sigma})^\mu_{\mathrm{hadron}}(g_{\mu\nu}+\chi_{\mu\nu})(1,\boldsymbol{\sigma})^\nu_{\mathrm{lepton}}\ ,
\label{currentsconnected}
\end{equation}
that shows the hadron and lepton current, connected by the $W$-boson propagator, we see that the effect of the tensor $\chi^{\mu\nu}$ is to connect the singlet and triplet parts of the lepton and hadron current in an unconventional way. For example the spacetime part of $\chi^{\mu\nu}$ connects the triplet of the hadron current to the singlet of the lepton current. This means that there can be a unit of spin change in the nucleus, while the lepton pair does not carry off the corresponding unit of angular momentum. This is the cause of the enhancement factor that occurs in some Lorentz-violating terms of the amplitude of transitions with $\Delta I \geq 2$. For a more elaborate discussion of this effect, see Refs.\ \refcite{lvbetadecay3} and\ \refcite{lvbetadecay2}.

\section{Bounds on Lorentz violation}
Using our results for a general $\beta$-decay transition we calculated the Lorentz-violating transition rate of a first forbidden transition with $\Delta I = 2$ and $\pi_i \pi_f = -1$, dependent on the emission direction of the outgoing $\beta$ particle. It is given by
\begin{eqnarray}
\frac{d\lambda}{d\Omega dE} & \propto & R^2 \Bigg\{ p^2 + q^2 \notag \\
&& + \ \frac{\alpha Z}{pR}\left[\frac{3}{10}\frac{p^3}{E}\left(\chi_r^{ij}\hat{p}^i\hat{p}^j-\tfrac{1}{3}\chi_r^{00}\right) \mp \frac{1}{2}p^2\tilde{\chi}_i^l \hat{p}^l \pm p^2 \chi_r^{l0}\hat{p}^l\right]\Bigg\}\ ,
\label{decayrspch2}
\end{eqnarray}
where $p$, $q$, $R$ are the $\beta$-particle momentum, the neutrino momentum and the nuclear radius respectively, $\tilde{\chi}^i = \epsilon^{ijk}\chi^{jk}$, Latin indices run over spatial components only, the subscripts $r,i$ denote real and imaginary parts respectively, and the upper (lower) sign corresponds to $\beta^-$ ($\beta^+$) decay. We see the relative enhancement factor $\alpha Z/ pR$ of the Lorentz-violating effects.
 
Using the expression in Eq.~\eqref{decayrspch2} we reanalyzed results from an experiment published in 1976 and described in Ref.~\refcite{newman}. The experiment was done using a 10 Ci strontium-90 source, which decays to yttrium-90 and subsequently to zirconium-90. The relevant decay is the one from $^{90}Y$ to $^{90}Z$, which is a first-forbidden transition with $\Delta I = 2$ and $\pi_i \pi_f = -1$. In the experiment the source and the detector were mounted on a turntable, which rotated with a frequency of 0.75 Hz. Meanwhile the current of electrons that originated from the source was measured. Using the data of 232 good 2-h runs, asymmetries were defined in terms of the emission direction of the electrons. These asymmetries were fitted to distributions in sidereal time, given by
\begin{equation}
\delta = a_0 + a_1 \sin(\omega t + \phi_1) + a_2 (2\omega t + \phi_2)\ .
\end{equation}
Bounds of the order of $10^{-8}$ were obtained on the constants $a_0$, $a_1$, and $a_2$. Using Eq.~\eqref{decayrspch2} we calculated the theoretical prediction for the asymmetries defined in Ref.\ \refcite{newman} and extracted the expressions for $a_0$, $a_1$, and $a_2$ in terms of $\chi^{\mu\nu}$. In this way we were able to put the following bounds at a $95\%$ confidence level\cite{lvbetadecay2}:
\begin{subequations}
\begin{eqnarray}
|2X_r^{30}-\tilde{X}_i^3| & < & 2 \times 10^{-8}\ , \\
|3 X_r^{33}- X_r^{00}| & < & 3 \times 10^{-6}\ , \\
\left[(X_r^{12}+X_r^{21})^2 + (X_r^{22}-X_r^{11})^2\right]^{1/2} & < & 1 \times 10^{-6}\ , \\
\left[(X_r^{13}+X_r^{31})^2 + (X_r^{23}+X_r^{32})^2\right]^{1/2} & < & 1 \times 10^{-6}\ , \\
\left[(2X_r^{20}-\tilde{X_i^2})^2+(2X_r^{10}-\tilde{X}_i^{1})^2\right]^{1/2} & < & 4 \times 10^{-8}\ ,
\end{eqnarray}
\end{subequations}
where the Lorentz-violating parameter $X^{\mu\nu}$ is now given in the standard Sun-centered intertial reference frame\cite{datatables}. These bounds are the best direct bounds on the parameter $\chi^{\mu\nu}$ of which we are aware.

\end{document}